# Ultrasensitive Anti-Stokes Luminescence Thermometry in Transition Metal Dichalcogenide Monolayers

Sharada Nagarkar[1*], Fahrettin Sarcan[1,2*], Elanur Hut[1,2], Emiliano R. Martins[3], Stuart A Cavill[1], Thomas F. Krauss[1] and Yue Wang[1*]

[1]School of Physics, Engineering and Technology, University of York, York, YO10 5DD, United Kingdom
[2]Department of Physics, Faculty of Science, Istanbul University, Vezneciler, Istanbul, 34134, Turkey
[3] São Carlos School of Engineering, Department of Electrical and Computer Engineering, University of São Paulo, São Carlos - SP, 13566-590, Brazil

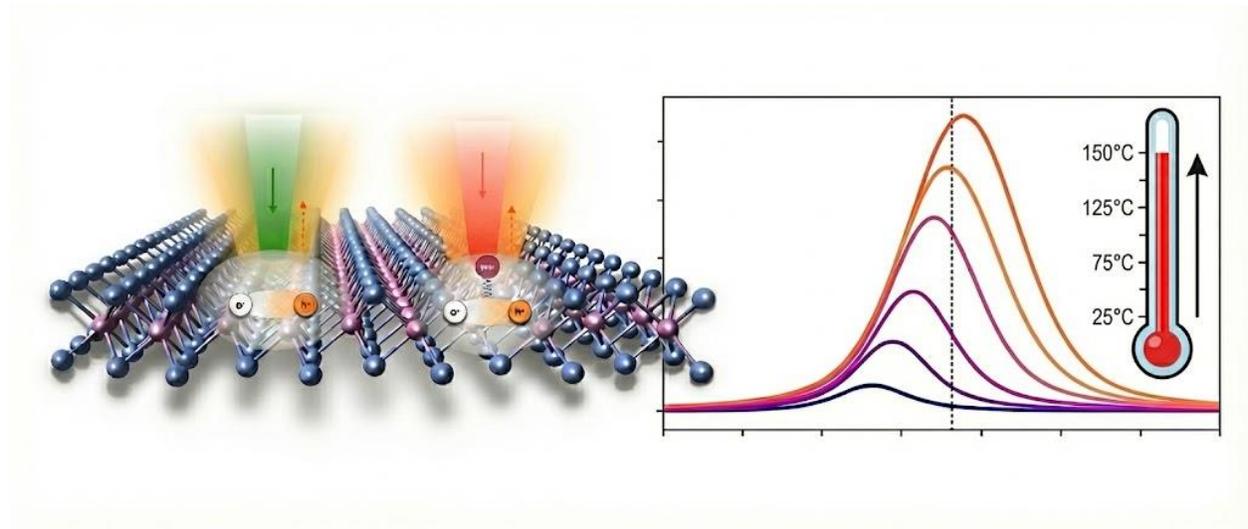

Abstract Figure

**Abstract**

Accurate temperature mapping at the nanoscale is a critical challenge in modern science and technology, as conventional methods fail at these dimensions. To address this challenge, we demonstrate a highly sensitive nanothermometer using anti-Stokes photoluminescence, also known as photoluminescence upconversion (UPL), in monolayer tungsten disulfide. Leveraging the direct band gap and strong exciton-phonon coupling in the two-dimensional monolayers, we achieve an exceptional relative sensitivity above 4% K$^{-1}$ across the 300 K to 425 K range, ranking it among the best-performing materials reported. A strong resonantly enhanced UPL is observed, confirming the central role of optical phonons in the upconversion mechanism. Furthermore, we introduce a new analytical model to quantitatively describe the UPL process, taking into account the interplay of phonon populations, bandgap narrowing, and substrate effects, which predicts resonant temperatures and provides a framework with broad applicability to any material exhibiting an anti-Stokes photoluminescence response. To demonstrate its use as a high-resolution optical thermometer, we map a 20 °C thermal gradient across a 20 µm long monolayer with a spatial resolution of 1 µm. With its high sensitivity, strong signal, and excellent reproducibility, our work establishes monolayer transition metal dichalcogenide as a leading platform for non-invasive thermal sensing in advanced microelectronic and biological systems.

1. **Introduction**

Accurate, non-invasive temperature measurements at the micro- and nanoscale are crucial for advancing technologies, ranging from next-generation electronics to cellular biology. Probing thermal dynamics in living cells, or mapping hotspots on microprocessors, demands techniques that offer high spatial resolution and are non-invasive. While non-contact infrared (IR) thermography appears to meet these criteria, it is fundamentally constrained by the spatial resolution inherent to the use of long-wavelength light, and the accuracy is often compromised by unknown and/or variable emissivity of the complex surfaces. In contrast, spectroscopy conducted at optical wavelengths provides a superior alternative as sub-micrometre regions can be probed[1–3].

Two techniques dominate the domain of optical spectroscopy-based thermometry. Firstly, Raman thermometry, which converts the ratio of the Stokes and anti-Stokes lines into temperature [4], and secondly, luminescence thermometry [5], which measures temperature-dependent changes in luminescence intensity, wavelength or lifetime. Although Raman-based thermometry is sensitive to phonon populations, it is constrained by the inherently weak Raman scattering cross-section, which results in low brightness and relative sensitivities well below 1% K$^{-1}$[6] . In luminescence thermometry, the most widely used probes have traditionally been rare-earth or transition-metal-doped materials, and more recently, quantum dots. While rare-earth-doped materials offer advantages such as narrow emission bands and wide operating ranges, they are often limited by low signal intensity, slow response times, and poor reproducibility[7,8]. Likewise, quantum dots provide high sensitivity, particularly for biological applications, but face significant challenges with reproducibility and thermal quenching of their luminescence at higher temperatures[9,10]. Therefore, a central challenge in this field is developing a luminescent probe that can simultaneously offer strong signal brightness and high thermal sensitivity and good reproducibility.

We suggest phonon-mediated photoluminescence upconversion luminescence (UPL) in 2D materials as a promising method to address this challenge. UPL is an intrinsically temperature-sensitive process, as its signal strength depends directly on the material's phonon population[11,12]. Materials that combine strong exciton-phonon coupling and high quantum efficiency are particularly suited, as their UPL signal is simultaneously bright and highly dependent on thermal changes. Two-dimensional transition metal dichalcogenides (2D TMDs) offer these properties; owing to quantum confinement, monolayer TMDs exhibit strong exciton-phonon coupling and high quantum yield[12–25], thus creating a luminescence signal that is both bright and highly responsive to temperature.

The brightness requirement, i.e strong UPL, has already been studied in several TMDs [12–14,26,27], but the mechanism of optical phonon coupling, leading to the thermal dependence of UPL and the thermometer function, is only poorly understood. Here, we fill this crucial gap by providing a comprehensive evaluation of monolayer $WS_2$ as a high-performance optical thermometer. In particular, we evaluate the key performance metric of relative thermal sensitivity ($S_r$), which is defined as the fractional change of temperature-dependent luminescence ($Q$) against temperature (Equation 1). Additionally, we identify the conditions for resonantly enhanced UPL, confirming the critical role of optical phonons in the upconversion process.

$$S_r = \frac{\partial Q/\partial T}{Q}$$

Eq. 1

## 2. Results and Discussion

### 2.1 Temperature-dependent PL, UPL and Raman spectroscopy

A monolayer $WS_2$ exfoliated on PDMS, shown in the optical micrograph in Fig. 1A, is placed on a sample stage with precise temperature control, see Fig. 1B. Figure 1C and 1D include schematic illustrations of the PL and the anti-Stokes PL, respectively. The $WS_2$ monolayer is confirmed by its room temperature direct bandgap emission peak at 620 nm (Fig. 1E), which agrees well with other studies[15,28,29]. Details on the curve fitting for PL and UPL spectra can be found in the Supporting Information.

UPL in monolayer $WS_2$ is facilitated by the presence of intermediate (in-gap) states, which lie energetically between the ground state and the bright exciton state. These in-gap states typically arise from intrinsic impurities, adsorbates, dark excitons, and local strain[13], representing a pathway for mediating the UPL process. The trailing edge of exciton and shallow in-gap trion states[30,31], which are observed as the broad and shallow low-energy shoulder peak in Fig. 1E, also contribute to the phonon-assisted UPL, in agreement with previous work[13]. Additionally, point defects, such as sulfur vacancies, can introduce deeper states located several hundred meV below the conduction band minimum, which can be responsible for multi-phonon-assisted upconversion emission[16,32,33].

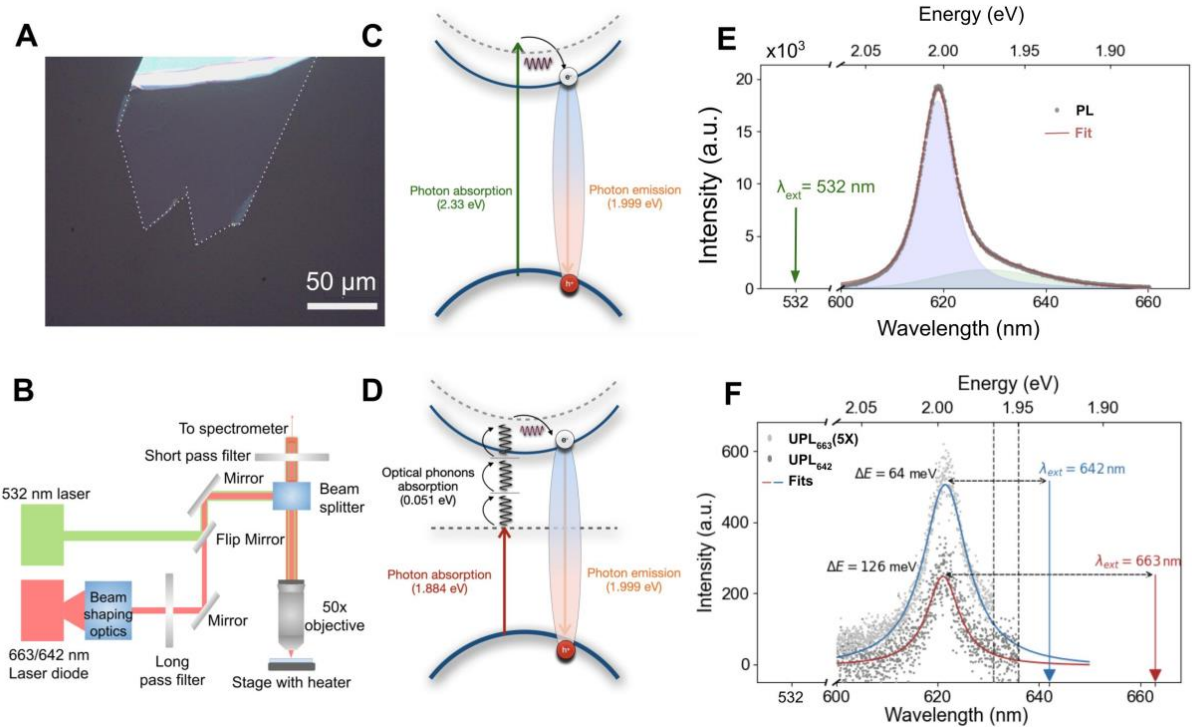

**Figure 1. PL and UPL (upconversion) from monolayer WS$_2$.** (A) Optical micrograph of a monolayer of WS$_2$ on a PDMS substrate, outlined by the dotted line, with a scale bar of 50 μm; (B) Schematic diagram of the micro-PL/UPL setup; (C) Schematic diagram of the PL (down conversion) process; (D) Schematic diagram of the UPL (up conversion) process; (E) PL spectrum (grey dot), measured with a 532 nm excitation at room temperature, fitted with a dual-Voigt model (red line); (F) UPL spectra (grey dot), measured with 642 nm and 663 nm excitation at room temperature, fitted with a single-Voigt model (solid line), respectively. The vertical dashed lines represent the cutoff wavelength of the short-pass filter for the corresponding laser.

Upconverted PL emission from the same monolayer was first measured with a 663 nm excitation laser. The laser power was maintained at 300 μW (power density of 1500 W/cm$^2$, same as the power density used for PL excitation). The energy difference between the pump and the emission peak is 126 meV ($\Delta E_1$ in Fig. 1F), which is provided by optical phonons. Secondly, we excite the UPL with a 642 nm laser, i.e. 1.93 eV, approximately 64 meV lower than the band gap ($\Delta E_2$ = 64 meV) to understand the impact of the energy difference; a smaller energy difference between pump and emission energies reduces the number of optical phonons required to bridge the gap. With the same pump intensity, we observe that the shorter excitation wavelength produces a significantly stronger UPL signal, shown in Fig. 1F. In absolute terms, the UPL signal is much weaker than the PL (2-3%), as is apparent from comparing Fig. 1E and 1F, yet the upconversion signal from monolayer WS$_2$ is strong and readily measurable.

We then measured the change in PL and UPL intensities with temperature, from room temperature up to 160°C (433 K). The PL intensity initially increases, reaches a maximum, and then decreases with increasing temperature above 110 °C (383 K), see Fig. 2A. The initial rise in PL can be attributed to the thermal delocalisation of excitons from shallow traps and thermally activated transfer from dark to bright states, as well as the desorption of surface contaminants, such as oxygen and water molecules at elevated temperatures[17,18]. Raising the temperature further inverts the trend, as higher temperatures promote the thermal dissociation of excitons and activate non-radiative recombination channels. In contrast, the UPL intensity rises steeply as the sample heats up across the entire temperature range, shown in Fig. 2B. Figures 2D and 2E plot the corresponding integrated PL and UPL intensities. For PL, the intensity exhibits an approximate fourfold increase from room temperature to 390 K, subsequently declining to 80% of its peak value around 420 K. For UPL, a continuous increase of nearly 1500-fold is observed from room temperature to 440 K.

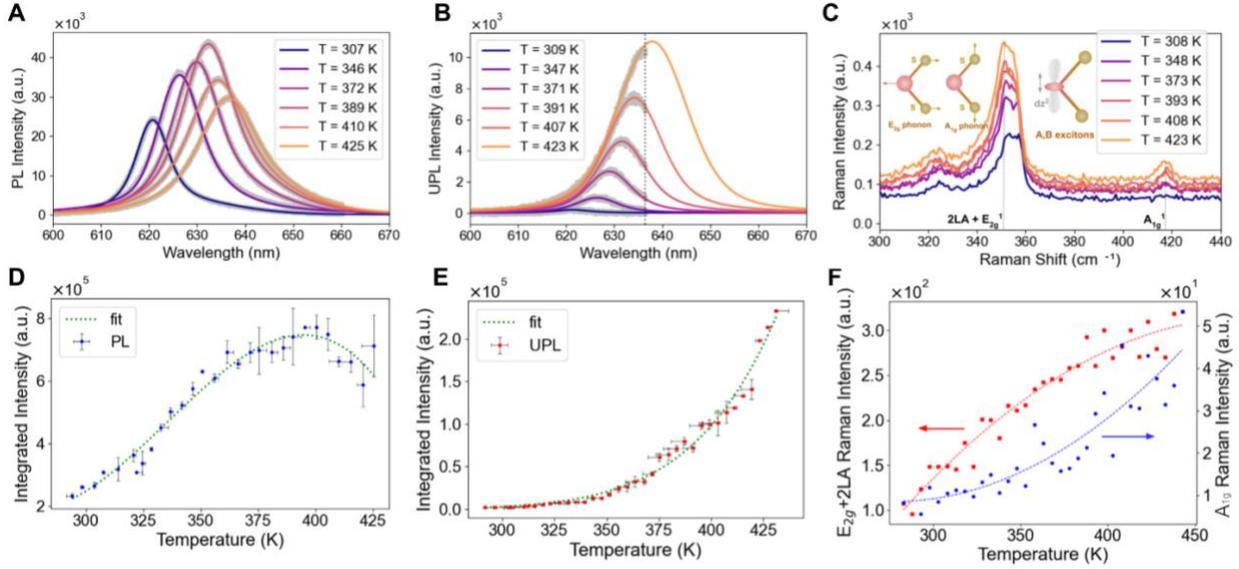

**Figure 2. Temperature-dependent PL, UPL and Raman measurements.** (A) PL spectra (raw data and fit) for six different temperatures, ranging from ambient to 425 K; (B) UPL spectra (raw data and fit) for six different temperatures, ranging from ambient to 423 K; (C) Raman spectra of monolayer $WS_2$ on a PDMS substrate, measured with 532 nm excitation, at various temperatures; Inset: Schematic drawing of $E_{2g}$ and $A_{1g}$ phonon modes along with the electronic orbitals of A and B exciton (recreated from ref[19]); (D) Integrated PL intensity vs temperature plot with a polynomial curve fit (guide to the eye); (E) Integrated UPL intensity vs temperature and a curve fit (green dashed line, based on the model discussed later in Section 2.3); (F) Temperature-dependent Raman intensity for $E_{2g}$ and 2LA modes (combined) and $A_{1g}$ modes, with a polynomial curve fit, respectively (red and blue dashed lines, guide to the eye).

The intensity increases in PL and UPL with increasing temperature are accompanied by a red shift in the peak wavelength of emission, as shown in Fig. 2, as a result of the well-known band gap reduction with temperature. PL and UPL both exhibit a similar wavelength shift, which follows the Varshni equation:

$$E_g(T) = E_0 - \frac{\alpha T^2}{T + \beta}$$

Eq. (2)

An analysis of the Varshni parameters for the monolayer is provided in SI Section 1. As the peak wavelength exhibits a continuous red shift with temperature, its value can be used to determine the sample's temperature. We measure its sensitivity as the fractional change in peak wavelength per degree, using Eq. 1. With the fitted curve, the calculated relative sensitivity is 0.02 % to 0.023 % $K^{-1}$ from ambient temperature to 160 °C (433 K). In section 2.3, we provide a comprehensive explanation of the UPL mechanism and an analytical model for the analysis of the relative sensitivity.

Analysing the UPL mechanism also requires consideration of the relevant phonon mode energies. We used Raman spectroscopy to determine the energy and population of these phonon modes in our monolayer sample. The Raman spectra, shown in Fig. 2C, feature the $A_{1g}$ optical phonon mode at 418 cm$^{-1}$ (i.e. 51.8 meV, shown in Fig. S3A) and the $E^1_{2g}$ optical phonon mode at approximately 353 cm$^{-1}$ (i.e. 43.6 meV), merged with the second-order longitudinal acoustic phonon, 2LA mode. The physical nature of these vibrations is illustrated schematically in the inset of Figure 2C (recreated from ref[19]). The bright A and B excitons are formed by direct-gap transitions at the K-point of the Brillouin zone in the monolayer, and are spatially localised almost entirely around the tungsten atoms[19]. The conduction band minimum at the K-point is therefore dominated by the out-of-plane $d_{z^2}$ orbital of the tungsten atoms. This results in a preferential and stronger phonon-exciton coupling between the A/B excitons with the out-of-plane $A_{1g}$ phonon mode than with the in-plane $E^1_{2g}$ modes, even though the $A_{1g}$ mode is weaker than the $E^1_{2g}$ mode in monolayer $WS_2$, see Fig. 2C. Therefore, the $A_{1g}$ phonon mode is critical for phonon-assisted UPL.

Given the $A_{1g}$ optical phonon's energy of 51.8 meV, bridging the energy gap between the 663 nm excitation and the UPL emission peak requires the participation of two to three phonons at room temperature. For 642 nm excitation, the required phonon number reduces to one to two, leading to a higher UPL yield. As the temperature is increased, the bandgap narrowing reduces the energy that must be supplied by phonons to facilitate the upconversion process. At the same time, the integrated intensity of the phonon peak also increases, see Fig. 2F, indicating that the number of optical and acoustic phonons both increases, governed by the Bose-Einstein distribution. Further analysis of the temperature-dependent Raman spectra is included in the Supporting Information Section 2. We can therefore conclude that as the temperature rises, two mechanisms work together to enhance the UPL intensity: the increasing population of phonons, most crucially, the $A_{1g}$ phonon, together with the temperature-induced narrowing of the band gap, which reduces the energy required for this process to occur.

## 2.2 Observation of resonantly enhanced UPL

As the nature of the substrate has an impact on the UPL process through phonon damping, we considered three different geometries, namely $WS_2$ monolayers on PDMS, on silicon nitride and in a suspended configuration. In all cases, the UPL intensity exhibited an exponential increase with temperature, irrespective of the substrate or excitation wavelength, see Fig. S1 and S2. However, a comparative analysis of these samples revealed a notable phenomenon.

In Figure 2E, the temperature-dependent intensity from the monolayer $WS_2$ on PDMS substrate exhibits a minor increase around 375 K. This anomaly is more pronounced and consistent in the suspended monolayer, highlighted in the boxes in Fig. 3A and 3B, occurring at approximately 365 K and 382 K, for 663 nm and 642 nm excitation, respectively. To investigate the origin of this anomaly, we analysed the energy difference (ΔE) between the incident photons and the emission peak. For the 663 nm excitation, the emission peak at this temperature is 628.3 nm (see Fig. 3C), resulting in ΔE = 103.3 meV. This energy is in close agreement with the total energy of two $A_{1g}$ phonons (2 x 51.8 meV). In the case of 642 nm excitation, the enhancement corresponds to a peak emission of 626.1 nm (see Fig. 3D), which yields ΔE ≈ 50.5 meV, closely matching the energy of a single $A_{1g}$ phonon[34,35]. Hence, the resonant enhancement in UPL is observed when the energy difference between the excitation energy and the emitted exciton energy is provided by an integer multiple of the $A_{1g}$ phonon energy, in other words, precisely matching the exciton's van Hove singularity and increasing the upconversion probability. This observation further confirms the crucial contribution of $A_{1g}$ optical phonons to the upconversion process.

Compared to the monolayers on PDMS and silicon nitride, the stronger resonance feature in the suspended monolayers reflects a stronger exciton-phonon coupling and the lack of alternative de-excitation channels. Our analytical model, discussed in the following section, can predict the resonant temperature, making it a broadly applicable tool for any material that exhibits phonon-assisted UPL.

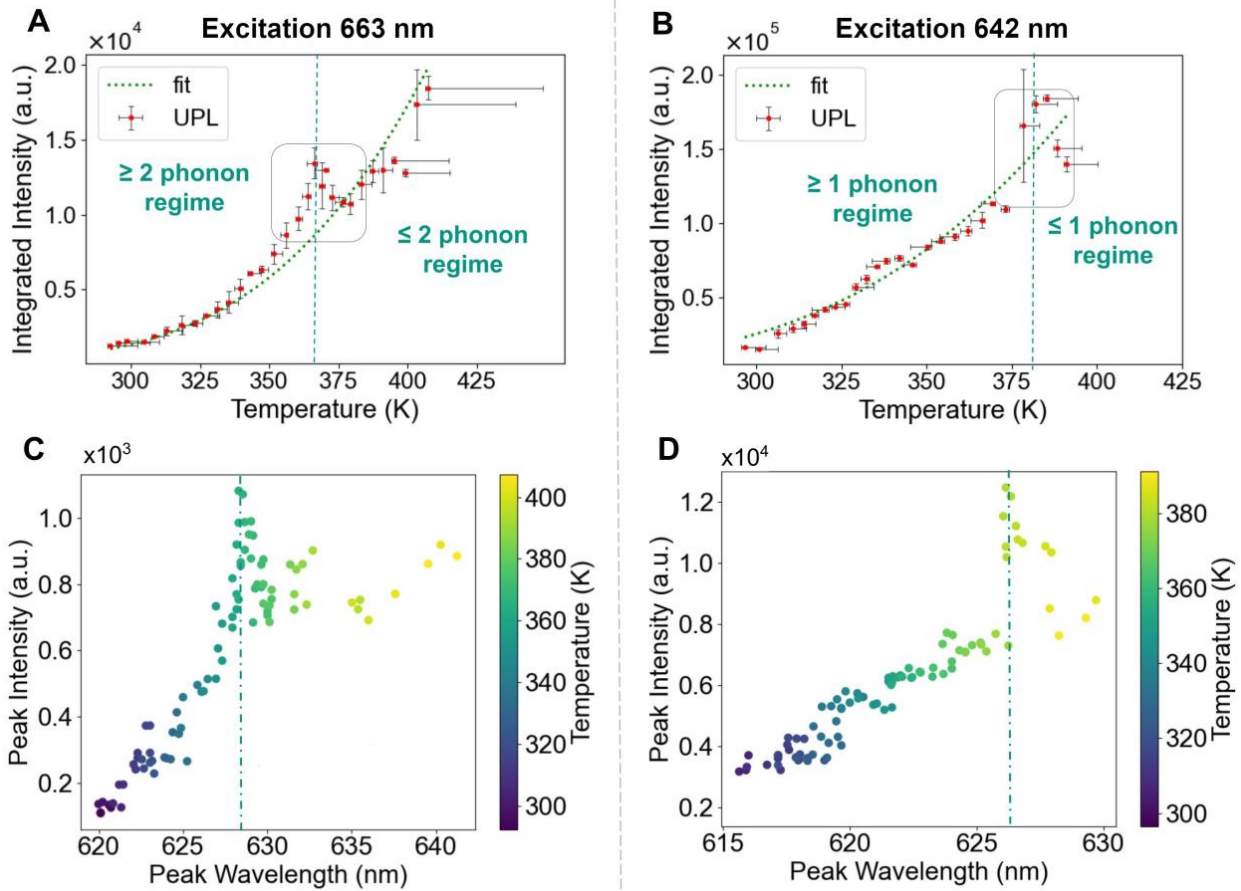

**Figure 3: Resonantly enhanced UPL in suspended monolayers.** (A) Integrated UPL intensity versus temperature for suspended monolayer $WS_2$ with a 663 nm excitation, with a resonant enhancement observed around 365 K; (B) Integrated UPL intensity versus temperature for a suspended monolayer $WS_2$ with a 642 nm excitation, with a resonant enhancement around 382 K. The green dashed lines are the fit using the model discussed in the next section. The grey boxes highlight the temperature windows of the resonantly enhanced UPL. (C) Temperature-dependent UPL peak wavelength and peak intensity with a 663 nm excitation, where a resonance is observed at a UPL peak wavelength of 628.3 nm, equivalently 365 K, and (D) with a 642 nm excitation, where a resonance is observed at a peak wavelength of 626.1 nm, equivalently 382 K.

## 2.3 Analytical model and relative sensitivity

Phonon-assisted UPL is often modelled with the Arrhenius equation in the literature, which reveals the activation energy[20]. The initial fit of our data showed a large activation energy for all substrates and incident wavelengths, ranging from 240 to 330 meV. This single-parameter fit is inconsistent and insufficient to fully account for the observed variations. Therefore, a comprehensive analytical model is required to clarify the distinct roles of the phonon population, Varshni bandgap shift, incident energy, and substrate effects.

As the $A_{1g}$ optical phonon that mediates the UPL process has a relatively flat dispersion[21], we can model its contribution using a single characteristic energy measured with Raman spectroscopy. Since UPL is a multiple phonon assisted anti-Stokes process, the probability of absorbing $n$ independent phonons is proportional to the Bose-Einstein phonon occupation number of the $A_{1g}$ optical phonon, $N(T)$, raised to the power of $n$, i.e. $N(T)^n$. Also, as temperature increases, the density of states at these wavelengths increases, see Supporting Information Section 4. We therefore fit the integrated upconversion intensity using the following expression:

$$I(T) = \eta\, \alpha(T)\, (c_s\, N(T))^{n(T)} \qquad \text{Eq. (3)}$$

where $\eta$ is a constant that accounts for both the photon collection efficiency of the setup and the PL quantum yield of the material, $\alpha(T)$ represents the absorption in the monolayer at the UPL excitation photon energy (details for fitting of $\alpha(T)$ are included in SI), and $c_s$, accounts for the exciton–phonon coupling coefficient.

$$N(T) = \frac{1}{\exp(E_{\text{phn}}/k_B T) - 1} \qquad \text{Eq. (4)}$$

$n$ or $n(T)$ is the number of $A_{1g}$ phonons at temperature T required to provide the energy difference between the incident/excitation photon and emitted photon, and is calculated as:

$$n(T) = \frac{E_g(T) - E_{\text{incident}}}{E_{\text{phn}}} \qquad \text{Eq. (5)}$$

Although $n$, representing the number of optical phonons required, should be an integer, our fitting allows a non-integer value of $n$, which accounts for the continuum of intermediate states and the thermal broadening of their distribution. The temperature-dependent bandgap $E_g(T)$ is obtained from the Varshni fit in Equation 2 [36,37]. As temperature increases, $N(T)$ grows while $n(T)$ falls, and the bandgap narrows. In our temperature range, the phonon occupation number and the coupling constant are both smaller than unity. These factors, combined with the increased density of states, result in the exponential temperature dependence observed in the UPL signal. Our model, described by Equations 3-5, accurately fits the temperature-dependent UPL intensity from different substrates. It should be noted that our model does not account for the resonant enhancement effect.

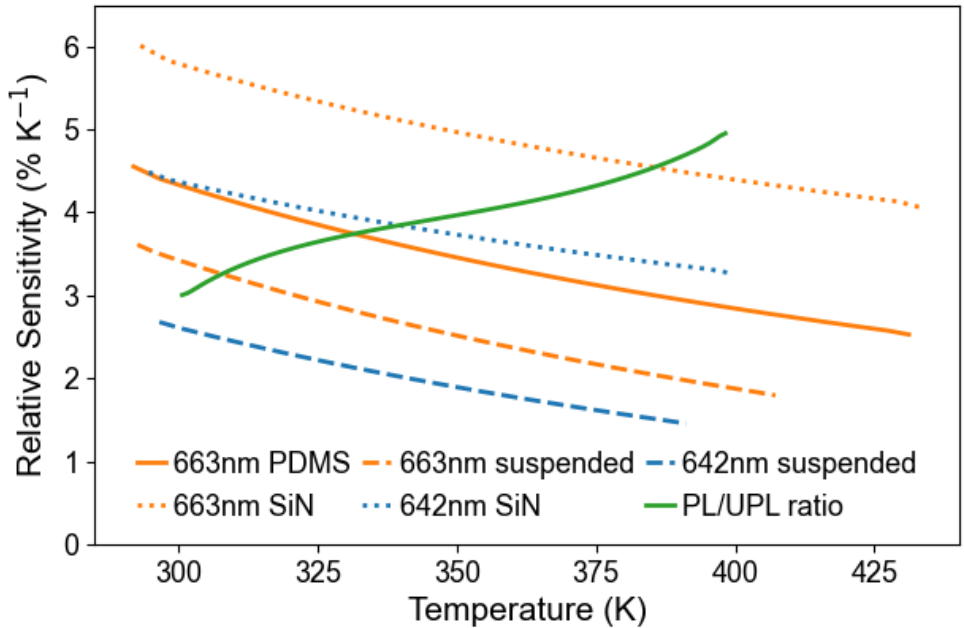

**Figure 4. Upconversion luminescence thermometry with different excitation laser wavelengths and substrates.** Relative sensitivities are derived for following experiments: 663 nm laser with monolayer on PDMS (orange solid line), 663 nm laser with suspended monolayer (orange dashed line); 663 nm laser with monolayer on silicon nitride (orange dotted line); 642 nm laser with suspended monolayer (blue dashed line); 642 nm laser with monolayer on silicon nitride (blue dotted line); and ratio of UPL to PL intensity with monolayer on PDMS (green solid line).

By fitting our model to the experimental data for all substrates and excitation wavelengths, we calculated the relative sensitivities for various scenarios according to Eq. 1, while excluding the resonance effect from the analysis. In Figure 4, we compared the relative sensitivities among monolayers on PDMS, on silicon nitride, and in suspended configurations, using intensity from PL, UPL and the ratio of two. Firstly, the non-ratio-based measurements reveal the highest relative sensitivity of above 4% $K^{-1}$ for UPL excited by the 663 nm laser with the monolayer on the silicon nitride substrate. For comparison, the relative sensitivity from the PL/UPL peak wavelength shift (Fig. S1) is 0.020 - 0.023 % $K^{-1}$ in the temperature range of 290 K to 433 K, so 200 times lower, while the Raman-derived sensitivities, calculated based on Fig. 2F, are 0.5 % $K^{-1}$ ($E_{2g}$ + 2LA mode) and 1 % $K^{-1}$ ($A_{1g}$ mode), still significantly lower. We also note that the suspended monolayer produces the brightest UPL intensity, but the lowest relative sensitivity. In fact, this observation is also supported by the model, which predicts an inverse relationship between the UPL brightness and the relative sensitivity, see Supporting Information Section 3. The monolayer on PDMS gives a thermal resolution of 95, 120, 108 and 79 mK, measured at 20, 40, 70 and 90 °C, respectively (see Supporting Information Section 5). In short, the higher the UPL intensity is, the better the thermal resolution is; however, the relative sensitivity decreases. Furthermore, the relative sensitivity of our UPL thermometry is several-fold greater near the resonance condition. This enhancement is due to the steep slope of the resonance peak, a characteristic that can be exploited for extremely sensitive thermometry.

We also implement the ratiometric thermometry method, which utilises the ratio of UPL to PL intensities from the monolayer. This approach is similar to those used in Raman and luminescent intensity ratio thermometry, serving as a self-referencing measurement that removes the need for separate calibration. Initially, up to 390 K, the sensitivity of this probe rises slowly because both the PL and UPL intensities increase, limiting the rate at which their ratio changes. However, this method gains a competitive advantage at elevated temperatures (beyond 390 K). In this higher temperature regime, the PL intensity decreases while the UPL intensity persistently escalates, a contrasting behaviour that causes the UPL/PL ratio to rise rapidly, resulting in an exceptionally high sensitivity.

Compared to other nanothermometer material platforms, such as lanthanide emitters, quantum materials, and perovskites, our sensitivity values are highly competitive. A summary of representative material platforms for nanothermometers is given in Table S3 in the Supporting Information Section 6. For example, Ge, Si vacancy in nanodiamonds achieved a sensitivity of 1.8% K$^{-1}$ at 300 K and 0.011% K$^{-1}$ at 388 K[22], while rare earth-doped double Perovskite 3.25% to 1.75% in the temperature range of 300 to 450K[23]. Additionally, the fabrication process of 2D-TMD thermometry is simple and reproducible. Monolayers can be reliably produced through mechanical exfoliation or chemical vapour deposition (CVD). This stands in contrast to the challenges encountered in achieving consistent doping concentrations in rare-earth materials or quantum dots, which are essential for other currently preferred thermometry materials. Additionally, the UPL-based thermometry technique offers notable advantages over Raman thermometry, primarily due to its significantly lower excitation power requirement[24,25,38]. It typically only requires 100s µW with short signal integration time, in contrast to the milliwatt to watt-level lasers commonly required for Raman spectroscopy. We also suggest that WS$_2$ monolayers can be directly integrated into existing microelectronic systems using a pick-and-place technique, enabling direct application in nano- and micro-scale devices without complex additional processing.

**2.4 Experimental demonstration of nanothermometry**

To demonstrate the practical utility of our method, we simulated and fabricated a chrome microheater on a silicon substrate, and subsequently positioned a monolayer adjacent to it, shown in Fig. 5A. An initial UPL scan, performed as a reference without activating the microheater, revealed a strain-induced redshift of 2 to 3 nm in UPL peak wavelength in regions near the edge of the flake. To decouple this strain effect from the thermal response, all subsequent measurements were restricted to the unstrained region between the two dashed lines in Fig. 5B. After activating the microheater, we performed a calibrated line scan, mapping a temperature gradient of approximately 20 °C across the length of the flake in 20 µm in steps of 1 µm (see Fig. 5C and 5D). Our spatial resolution was constrained by the 1 µm step size of our translation stage; however, the intrinsic spatial resolution of such a thermometer is much higher. We also performed a temporal measurement, where the UPL signal accurately resolved the heating and cooling cycles as the microheater was cycled on and off (see Fig. 5E). These experiments highlight that phonon-assisted UPL in TMD monolayers is a highly sensitive, high spatial resolution, non-invasive and easy to integrate technique for microscale thermometry.

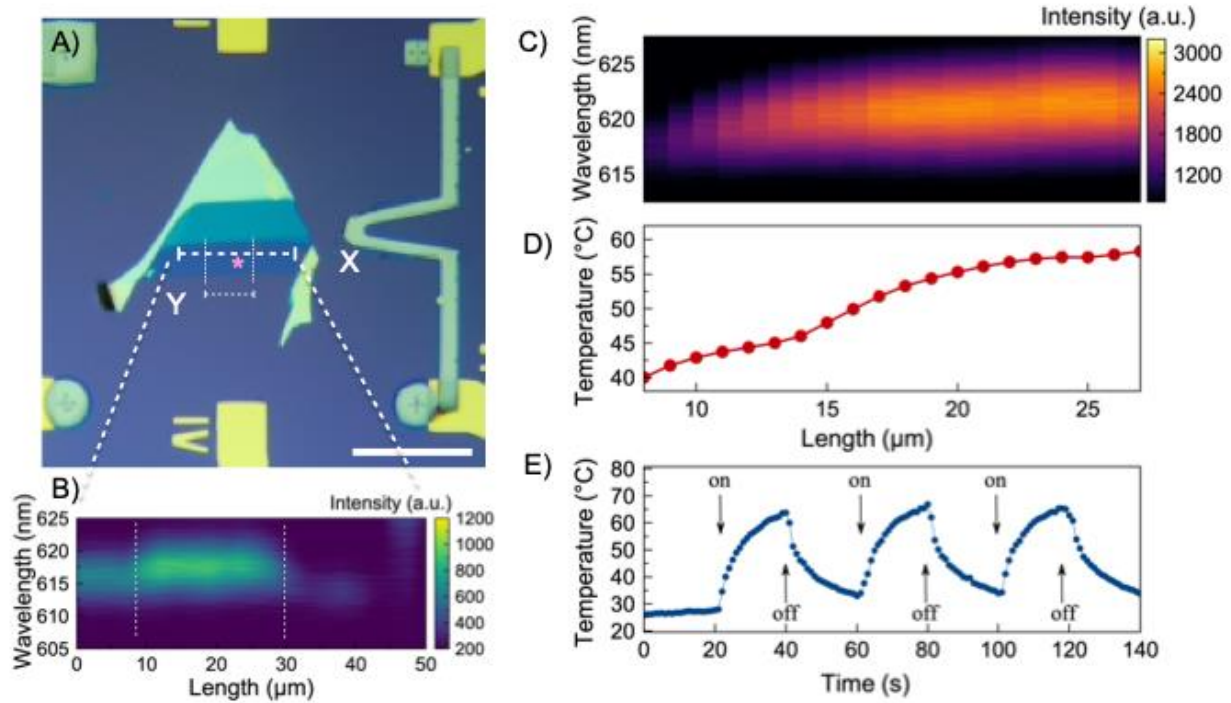

**Figure 5: Experimental demonstration of phonon-assisted UPL thermometry.** (A) optical microscopic image showing fabricated microheater (X) and a WS$_2$ flake (bulk and monolayer) (Y) placed near it, with a scale bar of 50 μm; (B) UPL measurements of the entire monolayer flake when the microheater is turned OFF, showing difference in peak wavelength as a function of position, caused by strain introduced in the sample fabrication process; (C) Spatial line map across the unstrained monolayer when the microheater is turned ON; (D) The resolved temperature gradience shows the area near the heater is approximately 20 °C hotter compared to the area that is 20 μm away from the heater; (E) A temporal map at a single point that is marked with a star in A, as a function of time. The microheater is periodically turned ON and OFF every 20 seconds.

**Conclusion**

In summary, we have demonstrated that monolayer WS$_2$ serves as an ultrasensitive, non-invasive platform for nanoscale luminescence thermometry. By exploiting optical phonon-mediated upconversion photoluminescence (UPL), our measurements on silicon nitride supported monolayers yield a peak relative sensitivity of above 4 % K$^{-1}$ across 300 to 425 K, highly competitive compared to state-of-the-art lanthanide-doped phosphors and quantum dot thermometry. Secondly, a resonant enhancement condition is observed when the energy difference between excitation and exciton emission corresponds to an integer multiple of the optical phonon energy, facilitating a precise match of the exciton's van Hove singularity; therefore, we can maximise the absorption and subsequent upconversion intensity. The analytical model we built upon Bose-Einstein phonon populations, band gap narrowing, and exciton–phonon coupling, effectively replicates the exponential temperature dependence of UPL intensity in experiments. Importantly, it quantifies the influence of phonon energy, incident photon energy, and the exciton-phonon coupling constant on both brightness and relative sensitivity. By spatially mapping a 20 °C thermal gradient with a 1 μm step size, we demonstrate that phonon-assisted UPL in TMD monolayers is a powerful and reliable technique for non-contact, optical thermal analysis at the micrometer scale. Exploring other 2D TMDs and heterostructures, employing a tuneable laser source, and operating around the resonance condition, could potentially broaden the operational range and further boost sensitivity. Additionally, enhancing the photonic density of states via nanoantenna[14] or photonic-crystal integration could significantly improve the UPL brightness and thermal sensitivity.

In short, monolayer $WS_2$ thermometry represents a promising contactless tool, offering a good balance of sensitivity and signal brightness, superior spatial resolution, and minimal invasiveness, enabling real-time nanoscale thermal monitoring.

**Methods**

**Temperature-dependent PL and UPL measurement of monolayer $WS_2$**
The monolayer $WS_2$ was mechanically exfoliated from a bulk crystal (2D Semiconductors) on a PDMS substrate. Photoluminescence was measured with a microPL setup with a CW laser (Novanta Photonics gem532) using a 50x objective. The emitted light was also collected through the same objective. The UPL was measured with laser diodes of two different wavelengths - 663 nm (L658P040 Thorlabs) and 642 nm (HL6364DG Thorlabs). In all measurements, incident power on the sample was kept the same, around 300 µW. In the measurements with the 663 nm laser, a 650 nm short-pass filter (Semrock) is used to block the incident laser from reaching the spectrometer. A 650 nm long-pass filter is placed after the diode laser to block any high-energy radiation from reaching the monolayer, ensuring that the observed emission is exclusively from PL upconversion. In the measurements with the 642 nm laser, we use a bandpass filter (FF01-642/10-25 Semrock) to block the high-energy radiation from the diode. All the spectroscopic measurements are performed with a high-resolution spectrometer (Acton Spectrapro 2750) coupled to an Andor CCD. The laser spot size on the sample is approximately 3 µm in diameter for excitation under a 50x objective.

To measure PL and UPL of monolayer $WS_2$ on a silicon nitride substrate, we transferred the exfoliated monolayer flake from $WS_2$ to a silicon nitride substrate via a micromanipulated transfer setup. The substrate consists of a 150 nm silicon nitride layer deposited on top of a silicon dioxide substrate. For the suspended monolayer samples, we patterned through a 150 nm silicon nitride membrane with holes, via lithographic patterning and dry etching, creating air holes with a diameter of 5 µm and then transferring the monolayers onto the air holes on the membrane.

For temperature-dependent PL and UPL measurements, the micro-PL setup is accomplished with a custom heating stage with a thermocouple and a PID controller. With the 663 nm excitation laser, the measurements span from 295 K to 423 K. For the 642 nm laser, we limit the measurements to 413 K, as above 413 K the red-shifted emission peak goes beyond the short pass filter cutoff and approaches the 642 nm excitation wavelength, resulting in a higher noise in the UPL data.

**Raman Spectroscopy**
A custom-built micro-Raman spectroscopy setup was employed to perform Raman measurements. A continuous-wave 532 nm laser (Novanta Photonics Gem532) was used for excitation, and the scattered signal was collected through a 100x objective (NA = 0.7). The laser beam was focused to a spot size of approximately 0.8 μm on the sample, which was mounted on a high-precision XYZ piezoelectric stage. The backscattered Raman signal was collected using the same objective. An 1800 grooves/mm diffraction grating was used in the spectrometer, providing a spectral resolution of approximately 0.7 cm$^{-1}$.

**Thermal mapping of monolayer WS$_2$**

Micro-luminescence thermometry test was performed to demonstrate the potential of these devices. A microheater was designed based on COMSOL Multiphysics simulation (see SI Section 7) to generate maximum local temperature gradient within a small area and subsequently fabricated on the substrate. The microheater was patterned by electron-beam lithography, and 50 nm of chromium was deposited using thermal evaporation, followed by a lift-off process. A mechanically exfoliated monolayer WS$_2$ flake was then deterministically transferred adjacent to the heater, and the device was mounted and wire-bonded onto a chip carrier. A constant current was applied to the heater to establish a temperature distribution, while the UPL signal of the WS$_2$ monolayer was monitored. All the UPL spectra were acquired in a kinetic series mode with an integration time of one second. Our spatial resolution is intrinsically limited to 1 μm by the minimum accurate displacement of the translation stage. For spatial mapping, a line scan was performed with a constant current of 10 mA applied across a pre-calibrated region with a uniform UPL signal. For temporal analysis, the excitation laser was focused on a single location, while the current supply to the microheater was periodically switched on and off every 20 seconds (7 mA, 20 V).

**Author Contributions**

S.N., T.F.K. and Y.W. conceived the idea. S.N. worked on sample preparation, temperature-dependent PL and UPL measurements. The analytical model was developed by S.N. with support from S.A.C. and E.R.M. F.S. and E.H. fabricated the suspended samples and performed the temperature-dependent Raman measurements. E.H., S.N., F.S. and Y.W. fabricated the microheater and performed the thermal mapping experiment. S.N., F.S. and Y.W. analysed the results and wrote the manuscript with contributions from all co-authors. Y.W. supervised and managed the project.

**Acknowledgments**

Y.W. and T.F.K. acknowledge support from the Engineering and Physical Sciences Research Council - EPSRC Grants EP/V047663/1. T.F.K. acknowledge funding from the Engineering and Physical Sciences Research Council – EPSRC EP/X037770/1 and Wellcome Trust (221349/Z/20/Z). F.S. acknowledges the Scientific Research Projects Coordination Unit of Istanbul University, project numbers FYL-2025-41674 and FBA-2025-41671.

**Data and code availability:**

The authors declare that all the data and code supporting the findings of this study are available within the article, or upon request from the corresponding author.

**Competing interests statement:**

The authors declare no competing interests.